%%%%%%%%%%%%%%%%%%%%%%%%%%%%%%%%%%%%%%%%%%%%%%%%%%%%%%%%%%%%%%%%%%%%%
% paper "Latticing quantum gravity",                                % 
% by R. Loll, begins here (tex-file)                                %
%%%%%%%%%%%%%%%%% fonts, definitions,etc.%%%%%%%%%%%%%%%%%%%%%%%%%%%% 

\input epsf.tex

\font\rmu=cmr10 scaled\magstephalf
\font\bfu=cmbx10 scaled\magstephalf

\font\it=cmti10 scaled \magstephalf
\font\bf=cmbx10 scaled\magstephalf
\rmu

\font\rmus=cmr8
\font\rmuss=cmr6
\font\mait=cmmi10 scaled\magstephalf
\font\maits=cmmi7 scaled\magstephalf
\font\maitss=cmmi7
\font\msyb=cmsy10 scaled\magstephalf
\font\msybs=cmsy8 scaled\magstephalf
\font\msybss=cmsy7
\font\bfus=cmbx7 scaled\magstephalf
\font\bfuss=cmbx7
\font\cmeq=cmex10 scaled\magstephalf

\textfont0=\rmu
\scriptfont0=\rmus
\scriptscriptfont0=\rmuss

\textfont1=\mait
\scriptfont1=\maits
\scriptscriptfont1=\maitss

\textfont2=\msyb
\scriptfont2=\msybs
\scriptscriptfont2=\msybss

\textfont3=\cmeq
\scriptfont3=\cmeq
\scriptscriptfont3=\cmeq

\newfam\bmufam  \textfont\bmufam=\bfu
      \scriptfont\bmufam=\bfus \scriptscriptfont\bmufam=\bfuss

\hsize=15.5cm
\vsize=21cm
\baselineskip=16pt   % Double spacing
\parskip=12pt plus  2pt minus 2pt

\def\semi{\bigcirc\kern-1em{s}\;}

\def\Q{{\mathchoice
{\setbox0=\hbox{$\displaystyle\rm Q$}\hbox{\raise 0.15\ht0\hbox to0pt
{\kern0.4\wd0\vrule height0.8\ht0\hss}\box0}}
{\setbox0=\hbox{$\textstyle\rm Q$}\hbox{\raise 0.15\ht0\hbox to0pt
{\kern0.4\wd0\vrule height0.8\ht0\hss}\box0}}
{\setbox0=\hbox{$\scriptstyle\rm Q$}\hbox{\raise 0.15\ht0\hbox to0pt
{\kern0.4\wd0\vrule height0.7\ht0\hss}\box0}}
{\setbox0=\hbox{$\scriptscriptstyle\rm Q$}\hbox{\raise 0.15\ht0\hbox to0pt
{\kern0.4\wd0\vrule height0.7\ht0\hss}\box0}}}}
\def\C{{\mathchoice
{\setbox0=\hbox{$\displaystyle\rm C$}\hbox{\hbox to0pt
{\kern0.4\wd0\vrule height0.9\ht0\hss}\box0}}
{\setbox0=\hbox{$\textstyle\rm C$}\hbox{\hbox to0pt
{\kern0.4\wd0\vrule height0.9\ht0\hss}\box0}}
{\setbox0=\hbox{$\scriptstyle\rm C$}\hbox{\hbox to0pt
{\kern0.4\wd0\vrule height0.9\ht0\hss}\box0}}
{\setbox0=\hbox{$\scriptscriptstyle\rm C$}\hbox{\hbox to0pt
{\kern0.4\wd0\vrule height0.9\ht0\hss}\box0}}}}

\font\fivesans=cmss10 at 4.61pt
\font\sevensans=cmss10 at 6.81pt
\font\tensans=cmss10
\newfam\sansfam
\textfont\sansfam=\tensans\scriptfont\sansfam=\sevensans\scriptscriptfont
\sansfam=\fivesans
\def\sans{\fam\sansfam\tensans}
\def\Z{{\mathchoice
{\hbox{$\sans\textstyle Z\kern-0.4em Z$}}
{\hbox{$\sans\textstyle Z\kern-0.4em Z$}}
{\hbox{$\sans\scriptstyle Z\kern-0.3em Z$}}
{\hbox{$\sans\scriptscriptstyle Z\kern-0.2em Z$}}}}

\newcount\foot
\foot=1
\def\note#1{\footnote{${}^{{}^{\number\foot}}$}{\ftn #1}\advance\foot by 1}

\def\frac#1#2{{#1\over #2}}
\def\text#1{\quad{\hbox{#1}}\quad}

\font\ch=cmbx12 scaled\magstephalf
\font\ftn=cmr8 scaled\magstephalf

\font\it=cmti10 scaled\magstephalf
\font\bf=cmbx10 scaled\magstephalf
\font\titch=cmbx12 scaled\magstep2
\font\titname=cmr10 scaled\magstep2
\font\titit=cmti10 scaled\magstep1
\font\titbf=cmbx10 scaled\magstep2

\nopagenumbers

%%%%%%%%%%%%%%%%%%% title page %%%%%%%%%%%%%%%%%%%%%%%%%%%%%%%%%%
\line{\hfil January 6, 1996}
\vskip3cm
\centerline{\titch LATTICING QUANTUM GRAVITY \note{To appear in the 
Proceedings of the 2nd Conference on Constrained Dynamics and Quantum
Gravity, Santa Margherita, Italy, 17-21 September 1996.}}
\vskip2.5cm
\centerline{\titname R. Loll}
\vskip.5cm
\centerline{\titit Max-Planck-Institut f\"ur Gravitationsphysik}
\vskip.2cm
\centerline{\titit Schlaatzweg 1}
\vskip.2cm
\centerline{\titit D-14473 Potsdam, Germany}
\vskip.3cm
\centerline{and}
\vskip.3cm
\centerline{\titit Erwin Schr\"odinger Institut}
\vskip.2cm
\centerline{\titit Boltzmanngasse 9}
\vskip.2cm
\centerline{\titit A-1090 Wien, Austria}

\vskip2cm
\centerline{\titbf Abstract}
\vskip0.2cm

I discuss some aspects of a lattice approach to canonical
quantum gravity in a connection formulation, discuss how it differs
from the continuum construction, and compare the spectra of 
geometric operators -- encoding information
about components of the spatial metric -- for some simple lattice quantum 
states.

\vfill\eject
\footline={\hss\tenrm\folio\hss}
\pageno=1
%%%%%%%%%%%%%%%%%%%%%%%%%%%%%%%%%%%%%%%%%%%%%%%%%%%%%%%%%%%%%%%%%%%%

\line{\ch 1 Discretizing gravity\hfil}

My contribution describes some aspects of an attempt to define a 
quantum theory of gravity in 3+1 dimensions, starting from a lattice
discretization of spatial 3-manifolds. This approach is 
complementary to other ones currently under study, most importantly,
the Regge calculus program and its variant, using so-called dynamical
triangulations. It differs from them in at least two aspects.
Firstly, its basic configuration variables are not discretized versions of 
the space- or space-time metric tensor, but of $su(2)$-valued 
connection one-forms. Secondly, in order to exploit the structural
resemblance with lattice gauge field theory, one best uses
Hamiltonian, and not path integral methods.  

\vskip1cm
\line{\bf 1.1 Setting\hfil}

The classical starting point is a reformulation of Einstein gravity
in terms of a phase space variable pair $(A_{a}^{i}(x),E^{a}_{i}(x))$ 
defined on a continuum 3-manifold $\Sigma$, where $a$ is a spatial 
and $i=1,2,3$ an adjoint $su(2)$-index. This is a real version of the 
well-known $su(2,C)$-valued Ashtekar variables (but still describing
Lorentzian, and not Euclidean gravity!). Using the 
real variables, one avoids the difficulty of having to impose quantum
reality conditions, but the Hamiltonian constraint
acquires a potential term which is not present in the complex 
formulation. This is functionally rather involved, but can probably  
still be handled [1]. The (doubly densitized, inverse) metric tensor is
a function of the momentum variables, $g^{ab}=E^{a}_{i}E^{bi}$.

In terms of the $(A,E)$-variables, Einstein gravity assumes the form 
of a Dirac constrained system, subject to a set of seven first-class 
constraints per space point, namely, three spatial diffeomorphism, one
Hamiltonian and three Gauss law constraints. 

In the lattice approach [2], one approximates $\Sigma$ by a lattice
$\Lambda$,
consisting of one-dimensio\-nal edges or links $l_{i}$ meeting
at vertices $n_{j}$. For simplicity, $\Lambda$ is chosen cubic, and 
all vertices are of valence six. The lattice analogues of
the Hamiltonian variables $(A,E)$ are a set of link-based variables
$(V,p)$ which however are {\it not} canonical. 
This comes about because the link analogue of the local {\it algebra}-valued 
connection $A(x)$ 
is the {\it group}-valued exponentiated integral of $A$, the link holonomy 
$V_{a}{}^{B}(l)$. Hence the configuration space associated with a single 
link is a copy of the compact group manifold $SU(2)$. 

The wave functions of the quantum theory are the square-integrable functions 
on the product over all links of the group $SU(2)$. The operators 
$\hat V_{a}{}^{B}(l)$ are 
represented by multiplication and the non-local link momenta $\hat p 
(l)$ can be identified with the left- and right-invariant vector fields on the 
$l$'th copy of the group. 

The kinematical structure described above is identical with the one
used in Hamiltonian lattice gauge theory. This setting is also well-suited 
for gravity, since the part of the constraints corresponding to 
internal gauge rotations is identical with those of Yang-Mills 
theory. One has two choices of dealing with the gauge 
constraints: one can either keep discretized versions of the quantum Gauss  
law constraints and eventually use them to project out physical, 
gauge-invariant wave functions, or go directly to a basis of 
gauge-invariant quantum states. We will follow the latter path, but this 
choice is not substantial. 

The elementary functions spanning the gauge-invariant Hilbert subspace
${\cal H}^{\rm inv}$ are the traces of holonomies of closed lattice
paths, obtained by multiplying together the corresponding link 
holonomies. An independent basis can be given in terms of so-called
spin-network states, where one assigns unitary irreducible representations of
$SU(2)$ (i.e. half-integer spins) to links and gauge-invariant
contractors to lattice vertices. However, the reader should be warned that in
explicit calculations one still has to worry about the presence of
zero-norm states, that exist in the form of Mandelstam constraints.
Equivalently, the choice of an independent set of states involves -- for 
fixed spin assignments -- a selection of {\it independent} contractors from
the entire set at each vertex $n$.
For each $n$, the spaces involved are finite-dimensional, 
but their dimension grows fast for increasing spins.   

\vskip1cm
\line{\bf 1.2 Lattice vs. continuum theory\hfil}

The lattice construction is in many 
aspects similar to the loop quantization program in the continuum, 
that also uses $SU(2)$-valued holonomies $U_{\gamma}[A]:={\rm P}\,
\exp\oint_{\gamma}A$ or their traces ${\rm Tr}\,U_{\gamma}[A]$ as the basic 
configuration variables. However, in order to avoid confusion, let us 
point out the main differences between the two formalisms. 

Graph or lattice configurations also appear in the continuum theory as
part of the specification of a quantum state. However, to obtain the 
entire Hilbert space of the kinematical quantum theory (i.e. before 
imposition of the Hamiltonian and diffeomorphism 
constraints), one has to consider states associated with {\it all possible} 
graphs. As a consequence, in order to specify a quantum state 
completely, one needs a) a graph $\gamma$, b) consistent non-vanishing 
spin assignments to all of its edges, and c) matching gauge-invariant 
contractors at all non-trivial vertices of $\gamma$. Of course, linear 
combinations of such states are also possible. The Hilbert space is,
loosely speaking, the space $L^{2}({\cal A}/{\cal G})$ of all square-integrable
functions on the space of gauge connections modulo gauge, which is an 
infinite-dimensional space.

By contrast, the configuration space for lattice gravity (for a finite
lattice) is {\it finite}-dimensional. Before considering gauge 
transformations, there are three degrees of freedom associated with 
each lattice link (parametrizing an element of $SU(2)$) times the 
number of links of the lattice. Furthermore, the lattice is {\it 
fixed}, i.e. all states and operators are defined on the same lattice. 
(Eventually, one wants to make the lattice bigger, 
in order to obtain a better approximation to the continuum theory. 
Still, the lattice operators never mix states associated with 
different lattices.) In order to specify a quantum state on the 
lattice, one needs a) consistent spin assignments to all of its edges
(vanishing spin is allowed, but does not imply that the underlying 
link ``vanishes''), and b) matching gauge-invariant contractors at all 
lattice vertices. 

A related important difference is that the continuum states depend on
graphs {\it imbedded} in $\Sigma$, whereas lattice states are 
based on a subset of lattice links, with the lattice $\Lambda$ 
itself not thought of as imbedded in an underlying manifold, but  
as a discrete approximation to $\Sigma$. As a result, in the 
continuum theory we can still define an action of the group of 
diffeomorphisms ${\rm Diff}(\Sigma)$ on states in a straightforward 
way. The lattice theory does not possess enough degrees of freedom to support 
such an action, and the most one can hope for is to define some kind 
of ``discrete version of ${\rm Diff}(\Sigma)$", that goes over to the 
usual continuum action in the limit as the lattice spacing $a$ is taken to 
zero. This is a non-trivial issue also in other discrete approaches to
quantum gravity.

Note that the appearance of one-dimensional ``loopy" excitations in 
the lattice theory is a consequence of the type of discretization we have 
chosen, and should not necessarily be considered fundamental, in the 
sense that as the continuum limit is approached, one may expect only genuine
three-dimensional properties of states and operators to be physically
important. The central assumption of the lattice construction is of 
course the existence of such a continuum limit. 

On the other hand, the 
fundamental assumption that leads to the continuum loop representation 
is that the Wilson loops ${\rm Tr}\,U_{\gamma}[A]$ become well-defined 
operators in the quantum theory. Classically, the
information contained in the ${\rm Tr}\,U_{\gamma}$ allows one to
reconstruct the space of {\it smooth} connections modulo gauge. 
Quantum-mechanically, the operators $\hat {\rm Tr}\,U_{\gamma}$ can 
be thought of as distributional excitations of the connection $A$ along
some loop or graph $\gamma$, and are therefore rather singular objects 
from a three-dimensional point of view. Nevertheless, well-behaved 
unitary representations of the classical algebra of the Wilson loops exist, and
it is exactly those that have been used in the loop quantization
approach. They do have some peculiar properties, for example,  
operator actions tend to be sensitive to certain topological characteristics
of quantum loop states, such as their number of edges or vertices, and
the way flux lines are arranged.
However, this is certainly not the only way one may set up a quantum
theory. It has been argued that it is physically more realistic 
to quantize configuration variables associated with three-dimensionally
smeared objects, like for instance tubes instead of loops.
Quantization of such an algebra is not likely to share all of the
features that have made the loop representations so attractive.

\vskip2cm

\line{\ch 2 Geometric operators\hfil}

In spite of the differences outlined in the previous section between the 
lattice and the continuum quantum theories, there obviously 
is a great structural resemblance between the two. This is in 
particular true for the action of certain geometric operators one may
construct in both settings, measuring volumes, areas, and lengths of 
spatial regions. It is not my aim here to discuss the construction 
and properties of these quantities in great detail, but rather to 
focus attention on a point that has not yet been addressed much in the
literature.    

The classical continuum expressions for volume, area and length are 
given purely as functions of the inverse dreibein variables $E$,

$$
{\cal V}({\cal R})=
\int_{\cal R} d^3x \sqrt{\frac{1}{3!}
\epsilon_{abc}\,\epsilon^{ijk} E^a_i E^b_j E^c_k},
$$
$$
{\cal A}({\cal S})=
\int_{\cal S} d^2x \sqrt{E^{3i} E^3_i},
$$
$$
{\cal L}({\cal C})=
\int_{\cal C} dx \sqrt{\frac{E^2_j E^{2j} E^3_k E^{3k}-
(E^{2j}E^3_j)^2}{\det E} },
$$

\noindent where ${\cal R}$ is a three-dimensional spatial region, ${\cal S}$ a
surface with unit normal in 3-direction, and ${\cal C}$ a curve dual 
to the 2-3-plane. As usual, their discretizations are not unique. We choose 
them as follows [3]:

$$
{\cal V}({\cal R}^{\rm latt})=\sum_{n\in{\cal R}^{\rm latt}} 
\sqrt{\frac{1}{3!} D(n)},
$$
$$
{\cal A}({\cal S}^{\rm latt})=\sum_{n\in{\cal S}^{\rm latt}}
\sqrt{\frac{1}{2} 
(p^+_i(n,\hat 3)p^{+i}(n,\hat 3)
+ p^-_i(n,\hat 3)p^{-i}(n,\hat 3))},
$$
$$
{\cal L}({\cal C}^{\rm latt})=\sum_{n\in{\cal C}^{\rm latt}}
\sqrt{\frac{3!}{D(n)}
( p_i(n,\hat 2)p^i(n,\hat 2) p_j(n,\hat 3)p^j(n,\hat 3) 
-(p_i(n,\hat 2)p^i(n,\hat 3))^2 ) },
$$

\noindent with $D(n)=\epsilon_{abc}\,\epsilon^{ijk}p_i(n,\hat a)p_j(n,\hat b)
p_k(n,\hat c)$, and the symmetrized link momenta $p_{i}(n,\hat a)$.

A nice property of the geometric lattice functions is that the expressions 
under the square roots can be represented by self-adjoint operators in 
the quantum theory, and therefore the operator square roots be defined in terms 
of the spectral resolutions. The spectra and eigenfunctions can be 
computed explicitly, by virtue of the fact that the quantum operators 
are defined purely in terms of the link momenta $\hat p$, which have a
particularly simple action on spin-network states. A further consequence 
is that the diagonalization of geometric operators can be performed 
independently at each lattice vertex (operators associated with 
different vertices commute), which vastly simplifies their
discussion. 

The complete spectrum of the area operator $\hat {\cal A}$ can be 
written down immediately, since $\hat {\cal A}$ is a function of
Laplacians only. The spectrum of the volume operator $\hat{\cal V}$ 
is only partially known, although general 
formulas for its matrix elements can be given. The spectrum of the 
length operator $\hat{\cal L}$ has not been studied yet. 

One reason for investigating the geometric operators is their 
simplicity, as compared to that of the Hamiltonian constraint. In 
addition, knowledge of the spectrum of the volume operator is vital
for constructing phase space functions depending in some way on
density factors of the form $\sqrt{\det g}\equiv\sqrt{\det E}$, for
example, the length function or the Hamiltonian. 

We will compare the spectra of these operators for some 
simple, explicit spin-network configurations. Since unit cells of the 
lattice can be regarded as the smallest building blocks of geometry,
one would certainly like to check whether the order of magnitude of 
the eigenvalues is comparable. To simplify matters further, we will 
consider maximally symmetric local lattice configurations, where no 
lattice direction is preferred, and concentrate on the volume and 
area operators. 

Recall that a local spin-network configuration around a vertex $n$ 
is determined by assigning half-integer spins $s_{i}$ or flux line
numbers $j_{i}=2s_{i}$ to each of the six incoming lattice edges,  
and a gauge-invariant contractor at $n$. We 
take all six $j_{i}$ equal, $j_{i}=j$, $j=1,2,\dots$. 
Fig.1 shows the length scales extracted from the 
area and (the non-negative) volume eigenvalues, i.e. the square and third root
respectively. For the volume eigenvalues, the degeneracy of the 
eigenspace is indicated. The area eigenspaces are maximally degenerate.
%\vfill\eject

\vskip1cm
\epsfysize=175pt\epsfbox{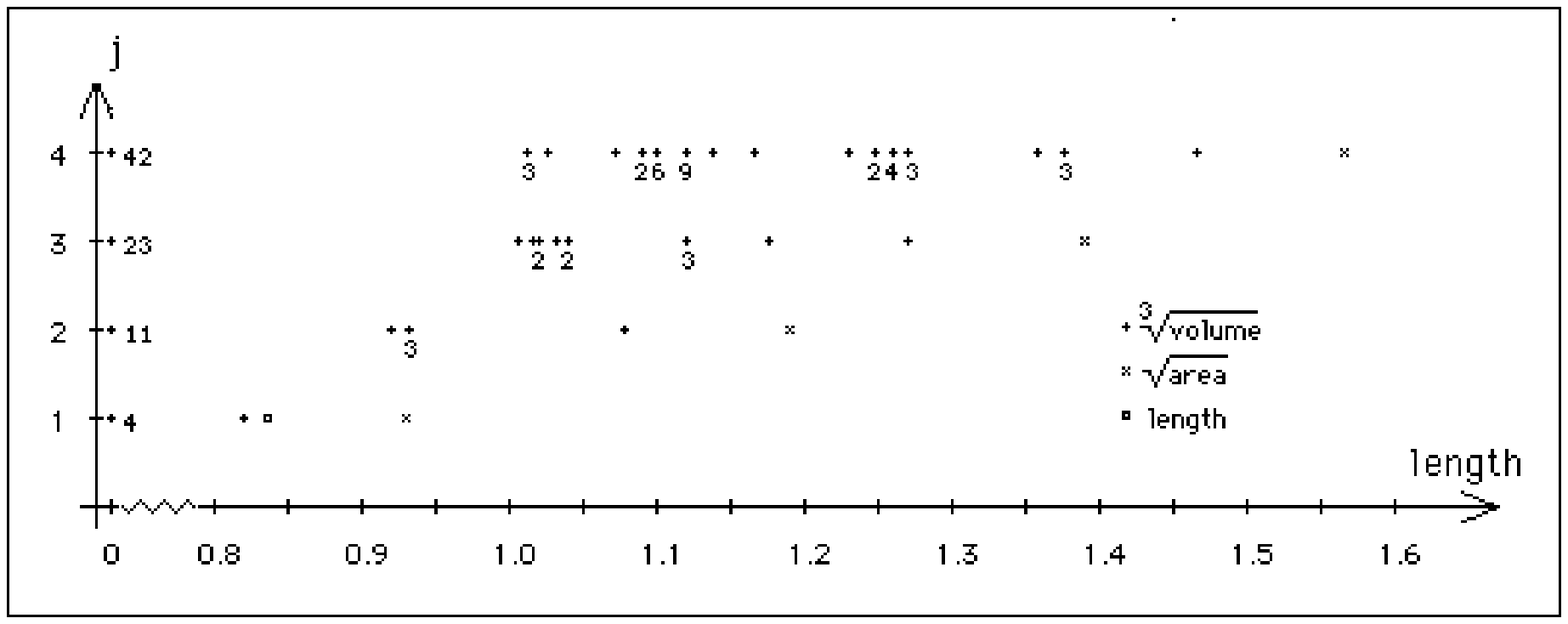}
\centerline{{\bf Fig.1} Eigenvalues of geometric lattice operators at a
given vertex $n$.}
\vskip1cm

Observe first that the length scales for given $j$ are roughly the same. 
This is also true for the single length eigenvalue calculated so far 
(it is that of the only positive-volume eigenstate for $j=1$). Beyond 
that, note that lengths obtained from the volume operator are
systematically smaller than those obtained from the areas. This 
indicates that one may encounter problems when attempting to construct a 
macroscopic flat, Euclidean geometry from these microstates, even if 
one uses eigenstates maximizing the volume for given $j$. It is 
possible that this effect goes away for larger $j$. 
Alternatively, this ``volume deficit'' may be an indication that generic 
local geometries have a small non-vanishing scalar curvature (I thank 
S. Carlip for this suggestion). 

Note also that the volume spectrum becomes more 
spread out with increasing $j$ and that there are {\it many}
zero-volume states. An important issue in quantum gravity is whether 
or not these states can or must be included in the Hilbert space.

\vskip2cm

\line{\ch References\hfil}

\item{[1]} R. Loll, Phys. Rev. D54 (1996) 5381; T. Thiemann, Phys. 
Lett. B380 (1996) 257.
\item{[2]} R. Loll, Nucl. Phys. B444 (1995) 619; Nucl. Phys. B460
(1996) 143.
\item{[3]} R. Loll, AEI {\it preprint}, Dec 1996.

\end

\hskip4.4cm\epsfxsize=150pt\epsfbox{s1dr.eps}
\vskip0.1cm
\centerline{{\bf Fig.1}}

\vskip0.3cm
\epsfysize=180pt\epsfbox{s11dr.eps}
\vskip0.1cm
\centerline{{\bf Fig.3}}